\documentclass[lettersize,journal,onecolumn]{IEEEtran}
\usepackage{amsmath,amssymb,amsfonts}
\usepackage{algorithmic}
\usepackage{algorithm}
\usepackage{array}
\usepackage[caption=false,font=normalsize,labelfont=sf,textfont=sf]{subfig}
\usepackage{textcomp}
\usepackage{stfloats}
\usepackage{url}
\usepackage{verbatim}
\usepackage{graphicx}
\usepackage{cite}

\usepackage{siunitx} 
\usepackage{multirow} 
\usepackage[version=4]{mhchem} 
\usepackage{xcolor}

\begin{document}



\title{Enhanced Read Resolution in Reconfigurable Memristive Synapses for Spiking Neural Networks}

\author{Hritom Das,~\IEEEmembership{Member,~IEEE}, Nishith N. Chakraborty,~\IEEEmembership{Member,~IEEE}, Catherine Schuman,~\IEEEmembership{Senior Member,~IEEE}, and Garrett S. Rose,~\IEEEmembership{Senior Member,~IEEE}
\thanks{Manuscript received June 23, 2023. This material includes research sponsored by the Air Force Research Laboratory under agreement number FA8750-21-1-1018. The U.S. Government may reproduce and distribute reprints for Governmental purposes, despite any copyright notation. The views and conclusions expressed herein are solely those of the authors and do not necessarily reflect the official policies or endorsements of the Air Force Research Laboratory or the U.S. Government. (Corresponding author: Hritom Das.)}
\thanks{Authors are with the Department of Electrical Engineering and Computer Science, University of Tennessee, Knoxville, TN 37996, USA (e-mail: hdas, nchakra1, cschuman, garose@utk.edu)}}

\markboth{}%
{Shell \MakeLowercase{\textit{et al.}}: A Sample Article Using IEEEtran.cls for IEEE Journals}


\maketitle

\begin{abstract}
Synapse is a key element of any neuromorphic computing system which is mostly constructed with memristor devices. A memristor is a two-terminal analog memory device. Memristive synapse suffers from various challenges such as forming at high voltage, \textit{SET}, \textit{RESET} failure, and \textit{READ} margin or resolution issue between two weights. Enhanced \textit{READ} resolution is very important to make a memristive synapse functionally reliable. Usually, the \textit{READ} resolution is very small for a memristive synapse with 4-bit data precision. This work considers a step-by-step analysis to enhance the \textit{READ} current resolution for a current-controlled memristor-based synapse. An empirical model is used to characterize the \ce{HfO2} based memristive device. $1^{st}$  and $2^{nd}$ stage device of our proposed synapse can be scaled to enhance the \textit{READ} current margin up to $\sim$ 4.3x and $\sim$ 21\% respectively. Moreover, \textit{READ} current resolution can be enhanced with run-time adaptation features such as \textit{READ} voltage scaling and body biasing. The \textit{READ} voltage scaling and body biasing can improve the \textit{READ} current resolution by about 46\% and 15\% respectively. 
TENNLab's neuromorphic computing framework is leveraged to evaluate the effect of \textit{READ} current resolution on  classification applications. Higher \textit{READ} current resolution shows better accuracy than lower resolution with different percentages of read noise scenarios.      

\end{abstract}


\section{Introduction}
\IEEEPARstart {T}{he} development of Artificial Neural Networks (ANNs) and Deep Neural Networks (DNNs) inspired by the remarkable information processing abilities of mammalian brains, while also achieving low power consumption and minimal latency. Due to their exceptional classification accuracy, DNNs are attracting considerable interest as the preferred classifier in numerous machine learning and computer vision applications \cite{intro1}. However, DNNs are typically executed on the von Neumann machines and are hence limited by the separation of memory and processing units, also known as the von Neumann bottleneck \cite{jetc}. Besides that, the extensive computational requirements, high power consumption, and memory bandwidth associated with DNNs make them unsuitable for mobile applications, where limitations in area and power are significant constraints \cite{intro2}.

In order to address the limitations in power and memory capacities of traditional computing architectures mentioned above, coupled with inspiration drawn from the efficiency of the biological nervous system, a novel concept of neuromorphic architectures has emerged. These architectures typically employ Spiking Neural Networks (SNNs), which aim to mimic the intricacies of the biological nervous system with greater fidelity by employing binary pulses as a means of communication. These architectures represent a distinct paradigm from the conventional von Neumann architectures in terms of the colocation of memory and processing unit,  demonstrating promising results in specific applications \cite{intro3,intro4,RC_glsvlsi}.
 Neuromorphic architectures not only offer better energy efficiency, but also promises parallel signal processing, fault tolerance, and reconfigurability. Furthermore, they can be implemented using diverse silicon-based technologies, large-scale architectures, and computational models of neural components \cite{mer,sensors,comp,adam_glsvlsi}.

 Neuromorphic systems also referred to as Neuroprocessors leverage the colocation of memory and processing units, where neurons act as the computational units, interconnected by synaptic memory elements. Synapses contain the weighted connections between neurons and can be implemented using digital \cite{adam} or analog \cite{glsvlsi2023,STDP} circuits. Nevertheless, incorporating a multitude of synapses presents several obstacles, including the efficient handling of storage space needed for weighted connections and the accommodation of diverse synaptic learning techniques that demand adaptable weight storage \cite{comp,Cruz}. Memristors are potential candidates to address these issues. Memristors have proven to be more compact and power efficient for synaptic implementation compared to SRAM and capacitor-based implementations of the same resolutions \cite{comp,brivio}. Memristive synapses have also been proved to have extended memory retention time \cite{brivio}.

 First postulated by Leon Chua, memristors are described as the fourth fundamental passive circuit element \cite{memr}. A memristor is a two-terminal device with analog memory properties that originate from its ability to switch resistances. When a voltage beyond a certain threshold is applied to its terminals, the resistance is modified. A memristor state is also non-volatile making it a promising candidate for weight storage. Due to their compatibility with CMOS technology and non-volatile properties, memristors are well-suited for analog computation \cite{ref_tab3}.

 In this work, we use a current-controlled synapse designed using a \ce{TiN TE\slash HfO2\slash TiN BE} memristor. To use this synapse, several operations need to be performed on the memristor, such as \textit{FORMING}, \textit{RESET}, \textit{SET}, and \textit{READ}. This memristor can vary its resistance in a few k$\Omega$ to over 150k$\Omega$ \cite{STDP,glsvlsi2023}. However, for our design, we exclusively use the low resistance states (LRS) of the memristors due to the improved reliability in this operating region, and to avoid a high degree of variability encountered near the high resistance states (HRS). The memristor can be programmed into different low-resistance states by precisely controlling the compliance current during \textit{SET} operation that overcomes the issues of variability and limited resolution \cite{r6}.

However, although the synapse is designed while taking the reliability concerns into consideration, the synapse is vulnerable to limited resolution due to the use of a narrow resistance range in the low-resistance regions \cite{r2}. This occurs when the current generated by the different resistance states is undifferentiable. This limitation can cause a reduction in the learning performance of SNNs \cite{r2,comp}. Overlapping synaptic currents also make the synapse to be susceptible to noise and process variation \cite{comp}. Another disadvantage of the small difference between the synaptic states is that it complicates the analog-to-digital converter (ADC) design significantly \cite{comp}. In order for the ADC to recognize different resistance states for digital conversion, the current output difference between the resistance states needs to be high enough for a compact power efficient design \cite{comp}. This work aims at improving the current resolution of the memristive synaptic circuit using several techniques.








The key contributions of this paper are as follows.  
\begin{enumerate}
    \item \textit{READ} current resolution of a current compliance memristive synapse is enhanced .
    \item \textit{READ} current resolution is enhanced with proper device scaling.
    \item \textit{READ} current resolution is made re-configurable at run time with \textit{READ} voltage scaling.
    \item \textit{READ} current resolution is adaptable at run time with body biasing.
    \item The TENNLab neuromorphic framework is utilized to observe the effect of \textit{READ} current resolution on SNNs. Higher \textit{READ} current resolution illustrates better accuracy with a lower possibility of a read error.      
    
\end{enumerate}

The rest of the paper is organized as follows. Section II briefly describes 
 a Verilog-A model, our proposed synapse with a hafnium oxide-based memristor device, and its \textit{READ} operation. Section III shows the proper device sizing to enhance the \textit{READ} current resolution. Section IV illustrates two techniques to make the \textit{READ} current resolution re-configurable at run time. Section V will evaluate the design performance based on different test cases. The next section (VI) exhibits the effect of the \textit{READ} current resolution or weight resolution on Spiking Neural Networks (SNNs). A detailed comparison with prior works is analyzed in Section VII. Finally, the paper is concluded in Section VIII with prospective future work. 

  \begin{figure}[]
            \centering
            \includegraphics[width=2in, height=2.5in]{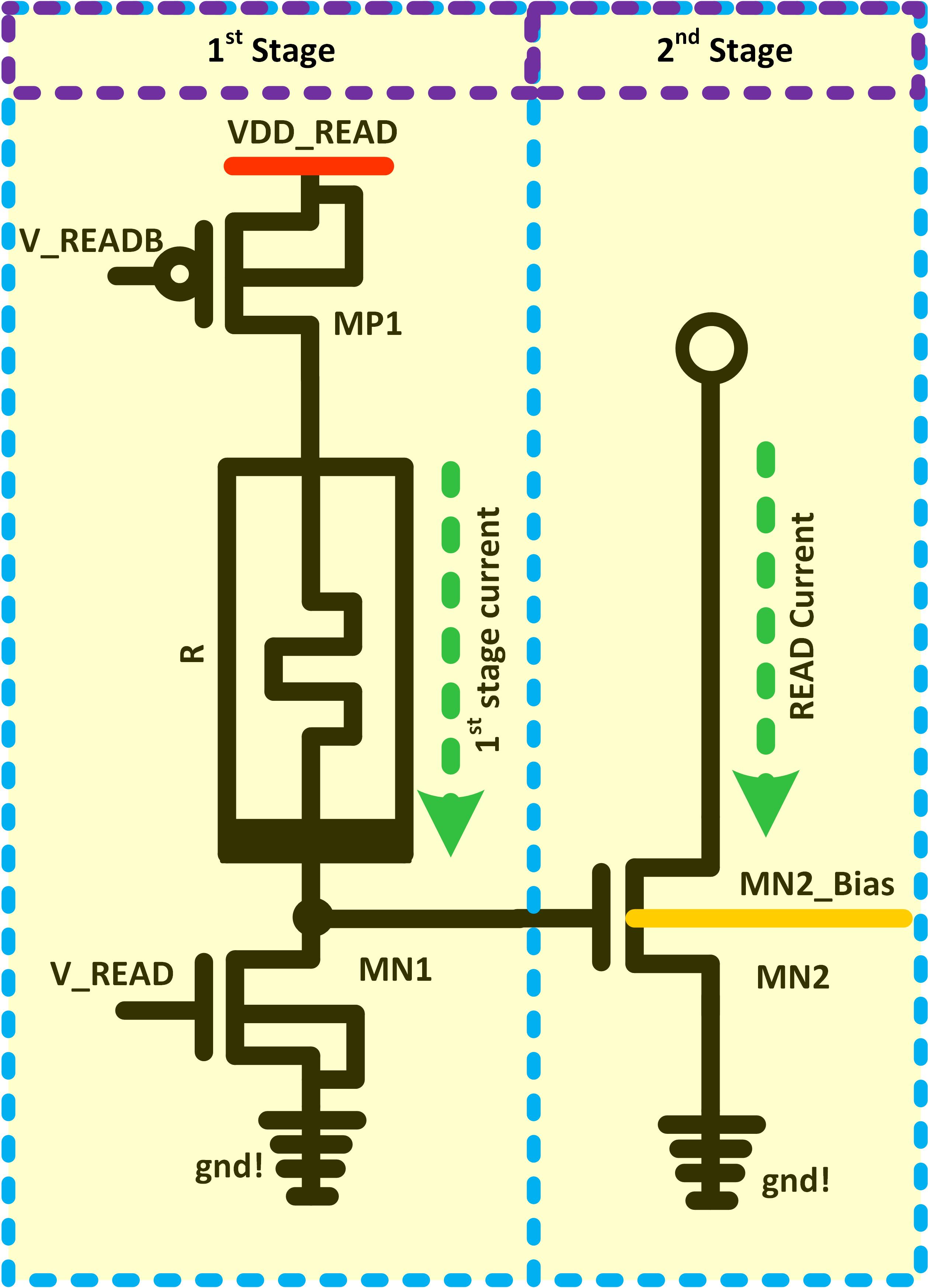}
            \caption{Memristor with \textit{READ} circuitry is illustrated. \textit{READ} operation requires 1-PMOS and 2-NMOS. \textit{READ} operation folded into two parts. $1^{st}$ stage current is generated with $M_{P1}$ and $M_{N1}$. This current initiates a voltage to operate $M_{N2}$ in the linear region. Finally, the \textit{READ} current will be sensed from the drain of $M_{N2}$. The body of the $M_{N2}$ is considered as a  dedicated signal to control the threshold of this device.}
            \label{fig:synapse_read}
        \end{figure}

\section{Current-Controlled Memristive Synapse}
\subsection{\ce{HfO2} Based Device Modeling}
A Verilog-A model is utilized to simulate the \ce{HfO2} based memristive devices\cite{model_glsvlsi}. In this model, mathematical equations are derived based on the memristance state and the required time to switch the states between HRS to LRS or LRS to HRS. The I-V characteristics of this device are taken under consideration to derive the empirical model of the memristor. The threshold voltage and switching time are two sets of important measured parameters for this model. There are some fitting constants in the model, which are utilized to fit the device's I-V characteristics to observe the simulation behavior as closely as measured data. The sigmoid window function is also considered to be incorporated with different patterns of switching time between HRS to LRS and LRS to HRS. This Verilog-A model is also capable of detecting the \textit{RESET} failure if the \textit{RESET} voltage is crossed its functional window. All mathematical equations, I-V curve, measured and, simulation details are available in our prior work\cite{model_glsvlsi}.            
\subsection{Proposed Synapse}
Memristors are widely used to construct brain-inspired synapses, which is the key element for neuromorphic computing. There are different synapse flavors based on the memristor's recipe (combination of materials). Various materials are utilized to build the memristors such as \ce{HfO2}, \ce{Ta2O5}, \ce{NbO2}, and so on. Our proposed architecture is designed using a \ce{TiN TE\slash HfO2\slash TiN BE} memristor. Fig. \ref{fig:synapse_read} shows the proposed synapse with \textit{READ} devices. However, the first step of this synapse is the one-time \textit{FORM} operation. Thick oxide transistors are used for this design to take care of high voltage around \SI{3.3}{\volt} for forming. A thick oxide transistor is also useful to reduce flicker noise.  Unformed memristors are usually exhibit resistance in the range of $\sim$\SI{8}{\mega\ohm} to $\sim$\SI{10}{\mega\ohm}.  After forming, the memristor's resistance level will be in few \SI{}{\kilo\ohm}. Hence, the synapse needs to \textit{RESET} to a higher resistance state (HRS), which is typically hundreds of \SI{}{\kilo\ohm}. Finally, our device is ready to \textit{SET} \slash write \slash program to a specific low resistance state (LRS) from a HRS. Due to less variability, the programming region is selected in the LRS region. The targeted LRS for this design is from \SI{5}{\kilo\ohm} to \SI{20}{\kilo\ohm}. As a result, we can utilize 4-bit precision with \SI{1}{\kilo\ohm} resistance resolution. This design needs a set of \textit{RESET}  and \textit{SET} to program in a new LRS value. After a successful \textit{SET} operation, the synapse is ready for a \textit{READ} operation.

Fig. \ref{fig:synapse_read} shows the proposed current-controlled synapse with \textit{READ} circuitry. $V\_READ$ and $V\_READB$ signals are utilized to access the memristor (R) during a \textit{READ} operation. $VDD\_READ$ and $V\_READ$ are \SI{1.2}{\volt} and \SI{0.6}{\volt} respectively during a \textit{READ} operation. Due to the \textit{READ} signal assertion, there will be a small $1^{st}$ stage current through the memristor. This current will create a voltage to operate transistor $M_{N2}$. Finally, a $READ$ $Current$ is sensed from the drain of the $M_{N2}$. The body of this $MOSFET$ is utilized as a separated signal to control the threshold voltage of $M_{N2}$. \textit{READ} is a very sensitive operation for memristor-based synapses. Especially, the \textit{READ} margin between two resistance levels (e.g. \SI{5}{\kilo\ohm} and \SI{6}{\kilo\ohm} ) needs to be good enough to read the data or weight properly. Most of the time the difference between the two resistance levels is few \SI{}{\nano\ampere} for low power design, which is very hard to sense properly. A research paper shows the \textit{READ} current between \SI{5}{\kilo\ohm} and \SI{6}{\kilo\ohm} is \SI{20}{\nano\ampere} \cite{r1}. A few techniques can be utilized to overcome this low \textit{READ} margin \slash resolution. All the techniques are explained below with proper analysis.  

\begin{figure}[]
            \centering
            \includegraphics[width=3in]{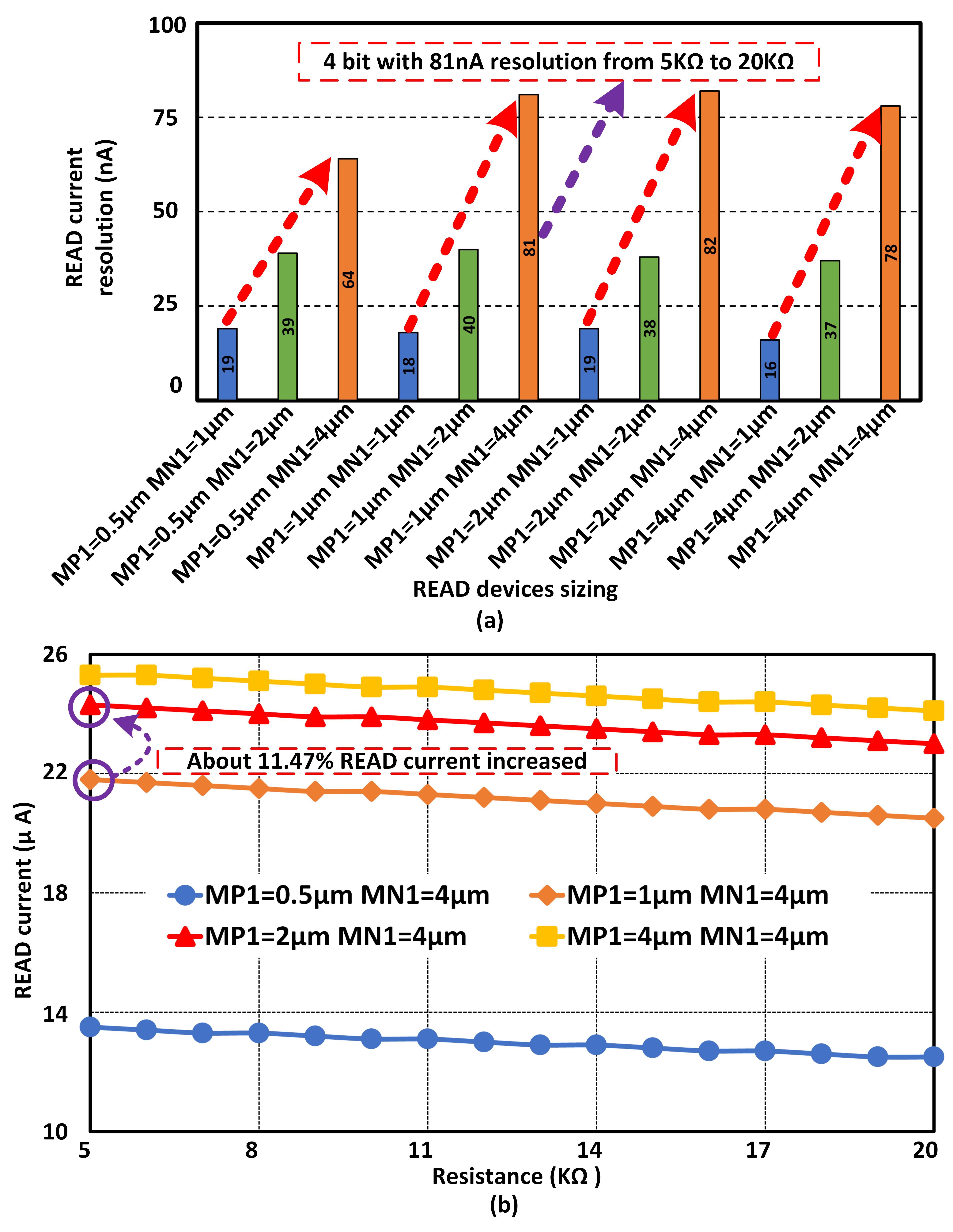}
            \caption{\textit{READ} simulation results are illustrated based on the sizing of $M_{P1}$ and $M_{N1}$. The length and width of $M_{N2}$ are fixed at \SI{0.5}{\micro\meter}. (a) $M_{P1}$ is varied from \SI{0.5}{\micro\meter} to \SI{4}{\micro\meter}. In addition,$M_{N1}$ is varied from \SI{1}{\micro\meter} to \SI{4}{\micro\meter}. Larger $M_{N1}$ shows a higher impact on the \textit{READ} current resolution. (b) shows the \textit{READ} current scale with different size of $M_{P1}$, when the width of the $M_{N1}$ is fixed at \SI{4}{\micro\meter}.}
            \label{fig:MP1_MN1}
        \end{figure}

\begin{figure}[]
            \centering
            \includegraphics[width=3in]{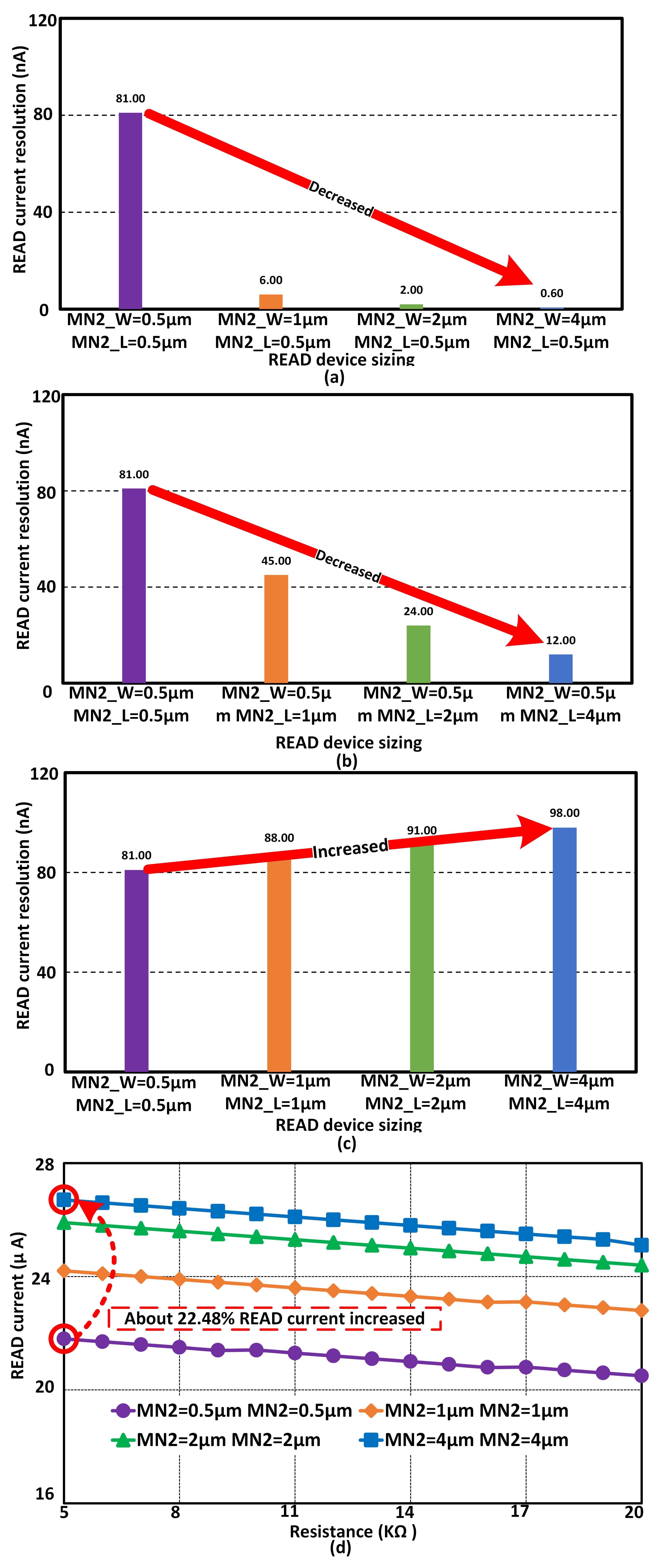}
            \caption{Cadence simulation results for \textit{READ} current resolutions with different width and length of $M_{N2}$ device. (a) shows the \textit{READ} current resolution when the length of the $M_{N2}$ is fixed at \SI{0.5}{\micro\meter} and the width is varied from \SI{0.5}{\micro\meter} to \SI{4}{\micro\meter}. The current resolution is drastically decreased with the increment of $M_{N2}$'s width. (b) shows the \textit{READ} current resolution when the width of the $M_{N2}$ is fixed at \SI{0.5}{\micro\meter} and the length is varied from \SI{0.5}{\micro\meter} to \SI{4}{\micro\meter}. The \textit{READ} current resolution is also decreased with the increment of length of $M_{N2}$. Finally (c) shows the \textit{READ} current resolution when the length and width of $M_{N2}$ change simultaneously. \textit{READ} current resolution is increased when the length and width of the $M_{N2}$ are increased at the same time. (d) shows the \textit{READ} current level with different $M_{N2}$ sizing. About 22.48\% \textit{READ} current overhead is observed to improve 21\% READ current resolution.}
            \label{fig:MN2}
        \end{figure}

\section {READ current resolution enhancement with proper device scaling}
At first, \textit{READ} device sizing is considered to observe the effect on \textit{READ} current resolution. A \SI{65}{\nano\meter} CMOS process is utilized to construct and conduct Cadence Spectre simulations. A Verilog-A model is utilized to characterize the \ce{HfO2} based memristor device \cite{model_glsvlsi}. Fig. \ref{fig:MP1_MN1}(a) shows effect on \textit{READ} current resolution with the sizing of $M_{P1}$ and $M_{N1}$. Here, the width and length of $M_{N2}$ are set at \SI{0.5}{\micro\meter}. In addition, the width of the $M_{P1}$ is varied from \SI{0.5}{\micro\meter} to \SI{4}{\micro\meter} and the width of the $M_{N1}$ is varied from \SI{1}{\micro\meter} to \SI{4}{\micro\meter}. In addition, the length of $M_{P1}$ and $M_{N1}$ (both are thick oxide transistor) are set at \SI{0.5}{\micro\meter}. When the width of both $M_{N1}$ and $M_{P1}$ is minimal, the \textit{READ} current resolution (one memristive level to another, e.g. \SI{5}{\kilo\ohm} to \SI{6}{\kilo\ohm}) is at least \SI{19}{\nano\ampere}. Due to an optimized read procedure, the \textit{READ} current shows very stable resolutions compared to prior work \cite{r1}. In this work, a regular pfet is utilized to control the read voltage at the drain of $M_N2$. Whereas a diode-connected pfet was connected in prior work.  Hence, the width of the $M_{P1}$ set at \SI{0.5}{\micro\meter} and the width of the $M_{N1}$ varies from \SI{2}{\micro\meter} to \SI{4}{\micro\meter}. The \textit{READ} current resolution is \SI{64}{\nano\ampere}, when the width of the $M_{N1}$ and $M_{P1}$ are  \SI{4}{\micro\meter} and \SI{0.5}{\micro\meter} respectively. The larger size of $M_{N1}$ and smaller size of $M_{P1}$ allow suitable gate voltage for the $M_{N2}$ to provide a high-resolution \textit{READ} current. Twelve different sizing combinations are observed for \textit{READ} current resolution. When the width of the  $M_{P1}$ and  $M_{N1}$ are  \SI{1}{\micro\meter} and \SI{4}{\micro\meter}, the \textit{READ} current resolution is about \SI{81}{\nano\ampere} for 4-bit data precision. According to the last test case, if the width of $M_{P1}$ increases significantly and its size becomes the same as $M_{N1}$, then the \textit{READ} current resolution does not show significant benefit on sizing. A better sizing combination is observed, when the $M_{N1}$ and $M_{P1}$ are not the same and $M_{P1}$ is smaller than $M_{N1}$. According to our sizing analysis, the \textit{READ} resolution provides best performance when the width of $M_{N1}$ and $M_{P1}$ are \SI{4}{\micro\meter} and \SI{1}{\micro\meter} respectively. 
Next, the effect of length and width of the $M_{N2}$ is observed, with the width of the $M_{N1}$ and $M_{P1}$ transistors are set at \SI{4}{\micro\meter} and \SI{1}{\micro\meter} respectively.

The length and width of $M_{N2}$ are varied to observe the effect on \textit{READ} current resolution. Fig. \ref{fig:MN2} (a) illustrates the \textit{READ} current resolutions, when the width of the device is varied from \SI{0.5}{\micro\meter} to \SI{4}{\micro\meter} and the length is fixed at \SI{0.5}{\micro\meter}. When the length and width of the device are set at a minimal size, the \textit{READ} current resolution is about \SI{81}{\nano\ampere}. The \textit{READ} current resolution is drastically decreased as the width of the $M_{N2}$ is scaled up. At the same time, the \textit{READ} current should be increased with $M_{N2}$ up-sizing. As a result, the first test case shows better \textit{READ} current resolutions with lower power consumption. 

Next, the width of the $M_{N2}$ is set at \SI{0.5}{\micro\meter} and the length is varied from \SI{0.5}{\micro\meter} to \SI{4}{\micro\meter}. The \textit{READ} current resolution is decreased with the increment of length. At this point, the overall \textit{READ} current will be decreased by up-sizing the length of $M_{N2}$, with higher latency. Finally, both length and width are increased simultaneously. Fig. \ref{fig:MN2} (c) shows the \textit{READ} current resolutions when both length and width are increased. \textit{READ} current resolutions are increased as the length and width are up-scaled at the same time. About 21\% improvement in \textit{READ} current resolutions can be achieved by up-sizing the length and width of $M_{N2}$ simultaneously. Fig. \ref{fig:MN2} (d) shows the \textit{READ} current level with different sizing combinations. \textit{READ} current resolution can be increased with the overhead of area and \textit{READ} current. Fig. \ref{fig:MN2} (d) shows, at \SI{5}{\kilo\ohm} memristive weight, the \textit{READ} current is \SI{21.8}{\micro\ampere} with minimal length and width of the $M_{N2}$. On the other hand, when both length and width of $M_{N2}$ are increased to \SI{4}{\micro\meter}, the \textit{READ} current is increased by 22.48\%. At the same time, the \textit{READ} current resolution is increased by 21\%. There is a clear trade-off between \textit{READ} current resolution and \textit{READ} current overhead. In addition, the overall design area is also influenced by a larger length of $M_N2$.

TABLE \ref{tab:sizing} shows the optimized sizing configuration to enhance the \textit{READ} current resolution. The width of $M_{P1}$ and $M_{N1}$ are set at \SI{1}{\micro\meter} and \SI{4}{\micro\meter} respectively. Both $MOSFET$'s length are fixed at \SI{0.5}{\micro\meter}. In addition, the length and width of $M_{N2}$ is considered as \SI{4}{\micro\meter}. The \textit{READ} current resolution is about \SI{98}{\nano\ampere} with this optimized sizing. In the next section, the \textit{READ} current resolution will be adapted dynamically with V\_READ and MN2\_Bias signals.          

\section{Reconfigurable READ current resolution }
\textit{READ} current resolution can be adapted at run time. Various applications can perform better at enhanced \textit{READ} current or weight resolution. To enhance the application's performance, a reconfigurable or run-time adaptation of \textit{READ} current resolution is proposed with different circuit techniques. Specifically, \textit{READ} voltage (V\_READ) scaling is a useful technique to influence the \textit{READ} current resolution at run time.       

\begin{table}[]
\centering
    \caption{Transistor scaling to enhance \textit{READ} current resolution}
\begin{tabular}{|c|c|c|}
\hline
Transistor name & Width (\SI{}{\micro\meter}) & Length (\SI{}{\micro\meter}) \\ \hline
MP1             & 1          & 0.5         \\ \hline
MN1             & 4          & 0.5         \\ \hline
MN2             & 4          & 4           \\ \hline
\end{tabular}%
\label{tab:sizing}
\end{table}

\begin{figure}[]
            \centering
            \includegraphics[width=3in]{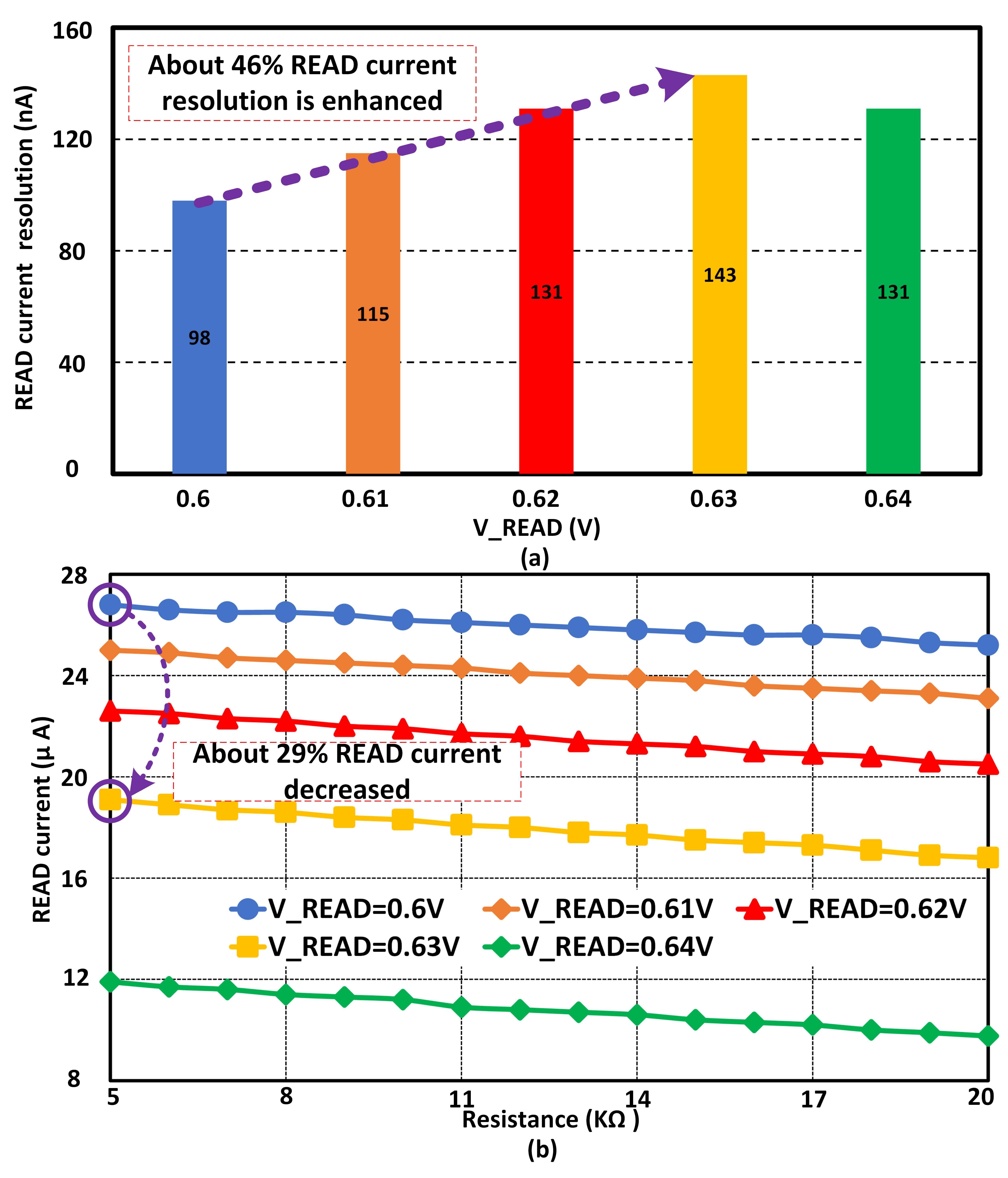}
            \caption{\textit{READ} voltage (V\_READ) has a significant effect on \textit{READ} current resolution. (a) shows the \textit{READ} current resolution at different V\_READ voltage. \textit{READ} current resolution is increased with the increment of the gate voltage of $M_{N1}$. After a certain level of gate voltage increment, the \textit{READ} resolution starts decreasing. (b) exhibits the \textit{READ} current level at different \textit{READ} voltages. As we increase the \textit{READ} voltage at the gate of $M_{N1}$, the \textit{READ} current level starts decreasing. Due to a weak turn-on of $M_{N2}$, the \textit{READ} current level is decreased. Here is an interesting thing to notice, as we increase the gate voltage of $M_{N1}$ the overall \textit{READ} current level is decreased. As a result, the \textit{READ} current resolution is increased with overall \textit{READ} current optimization.}
            \label{fig:READ voltage}
        \end{figure}

\subsection{READ Voltage Scaling}
Fig. \ref{fig:synapse_read} shows V\_READ is the gate voltage of $M_{N1}$. Initially, the applied amplitude of this signal was \SI{0.6}{\volt}. Fig. \ref{fig:READ voltage} (a) shows the \textit{READ} current resolution at \SI{0.6}{\volt} is \SI{98}{\nano\ampere} for 4-bit data. Here, the optimized sizing is utilized from TABLE \ref{tab:sizing}. The \textit{READ} current resolution is about \SI{143}{\nano\ampere} at \SI{0.63}{\volt} as V\_READ. About 46\% of \textit{READ} current resolution can be enhanced with V\_READ scaling at run time. If the V\_READ is increased more than \SI{0.63}{\volt} for this particular sizing or configuration, then the resolution is decreased. A stronger turn-on of $M_{N1}$ influences the final \textit{READ} current negatively. As a result, the resolution is decreased with excessive V\_READ.  Fig. \ref{fig:READ voltage} (b) shows \textit{READ} current with different gate voltages of $M_{N1}$. When the \textit{READ} voltage is \SI{0.6}{\volt}, the \textit{READ} current is about  \SI{26.8}{\micro\ampere} at \SI{5}{\kilo\ohm}. On the other hand, if the \textit{READ} voltage is scaled up to \SI{0.63}{\volt}, the \textit{READ} current will be decreased to about \SI{19.1}{\micro\ampere}. About 29\% \textit{READ} current can be optimized when the \textit{READ} voltage is scaled from \SI{0.6}{\volt} to \SI{0.63}{\volt}. \textit{READ} current shows a significantly lower value at \SI{0.64}{\volt}. However, at \SI{0.64}{\volt} the \textit{READ} current resolution is reduced significantly compared to the value at \SI{0.63}{\volt}. In addition, at \SI{0.64}{\volt} the std. dev. of \textit{READ} current is $\sim$ 7\% higher than std. dev. at \SI{0.63}{\volt}. Thus, the \textit{READ} current at \SI{0.63}{\volt} is more reliable than at \SI{0.64}{\volt}. According to our design optimization, the \textit{READ} current resolution is higher at \SI{0.63}{\volt} among all the \textit{READ} voltages. In addition, the \textit{READ} current resolution can be varied from \SI{98}{\nano\ampere} to \SI{143}{\nano\ampere} with V\_READ signal scaling. In the next sub-section, another device technique is utilized to enhance the \textit{READ} current resolution at run time.

\begin{figure}[]
            \centering
            \includegraphics[width=3in]{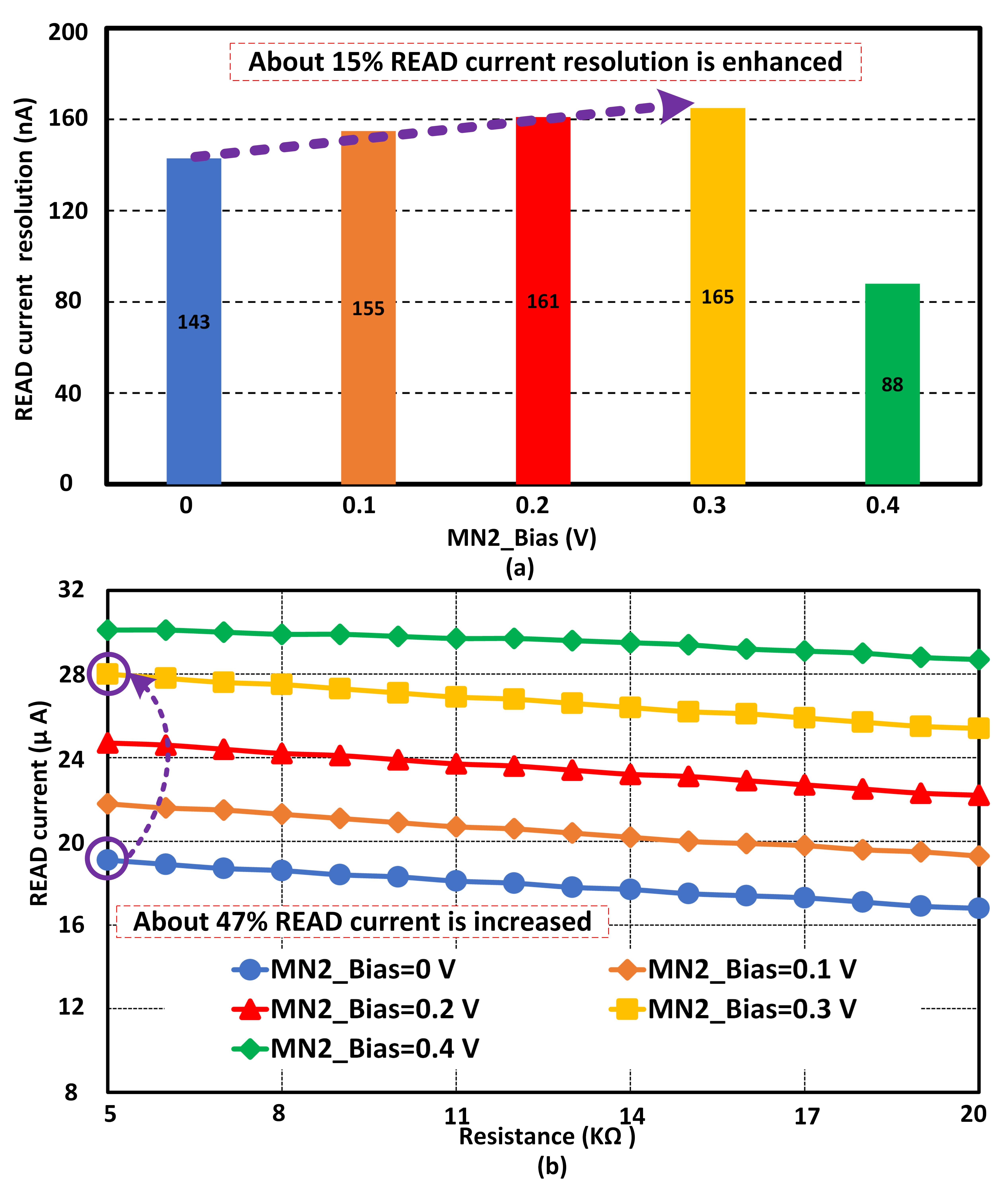}
            \caption{\textit{READ} device $M_{N2}$ plays an important role to make \textit{READ} current resolution adaptable at run time. The body of $M_{N2}$ is scaled to enhance the resolution with \textit{READ} power overhead. (a) shows the body biasing effect on \textit{READ} current resolution. About 15\% resolution can be enhanced with body biasing. (b) illustrates the \textit{READ} current level at different biasing voltages. The \textit{READ} current is increased with the increment of body biasing of $M_{N2}$.}
            \label{fig:biasing}
        \end{figure}
        
\subsection{ READ Device $M_{N2}$ Biasing}

\textit{READ} device $M_{N2}$ biasing is another technique to manage synaptic \textit{READ} current at run time. Here the \textit{READ} voltage V\_READ is set at \SI{0.63}{\volt} with optimized sizing. Fig. \ref{fig:biasing} (a) shows the \textit{READ} current resolution at different bias voltages at the body of $M_{N2}$. As we know body biasing can change the threshold voltage of a $MOSFET$ to control the current flowing through it. Here a positive body voltage is applied on $M_{N2}$ to improve the \textit{READ} current resolution. When the body voltage is \SI{0}{\volt}, the \textit{READ} current resolution of our proposed synapse is about \SI{143}{\nano\ampere}. If the body voltage is scaled up to \SI{0.3}{\volt}, then the \textit{READ} current resolution is further up to 15\%. In addition, if the body biasing voltage is increased more, then a reverse phenomenon is observed with \textit{READ} current resolution. Due to the channel effect of the NMOS, the reverse phenomenon is observed with higher body bias voltage.  At \SI{0.4}{\volt} the \textit{READ} current resolution is decreased to  \SI{88}{\nano\ampere}. As a result \SI{0.3}{\volt} is an optimized body bias voltage for this design scenario. Fig. \ref{fig:biasing} (b) exhibits the \textit{READ} current of our synapse at different body biasing scenarios. Overall \textit{READ} current is increased with the increment of body biasing voltage. About 47\% \textit{READ} current is increased with body biasing to enhance the  \textit{READ} current resolution at run time. Finally, it can be said that the \textit{READ} current resolution of our proposed synapse can be adapted at run time with different body biasing voltages.



\section{READ current resolution and design performance evaluation}

TABLE \ref{tab:table2} shows an evaluation of device sizing, \textit{READ} current resolution, and \textit{READ} power. There are five test cases considered for the evaluation. The first test case is constructed with base device sizing. Here, the V\_READ and body bias (MN2\_Bias) are \SI{0.6}{\volt} and \SI{0}{\volt} respectively for first three test cases. The first test case shows \SI{19}{\nano\ampere} \textit{READ} current resolution with \SI{18.87}{\micro\watt} as a max \textit{READ} power. Here, both stages' power ($1^{st}$ and $2^{nd}$) are considered for the max \textit{READ} power. The second test case is the $1^{st}$ stage device sizing. Here the \textit{READ} current resolution is \SI{81}{\nano\ampere} with 4-bit precision which is about 4.3x higher than the base sizing test case. Only 5\% of power overhead is observed compared to the base test case. The next test case is focused on $2^{nd}$ stage device sizing with $1^{st}$ stage sizing. Here, the \textit{READ} current resolution is \SI{98}{\nano\ampere}, which is 5.16x more improved than the base test case (Base sizing). Here, the \textit{READ} power overhead is only 0.16\% compared to the base test case. The fourth test case is to increase the V\_READ to \SI{0.63}{\volt} from \SI{0.6}{\volt}. In this scenario, the \textit{READ} current resolution is \SI{143}{\nano\ampere}, which is 7.53\% enhanced than the base test case. In addition, the max \textit{READ} power shows 1.43\% overhead compared to the base test case. Finally, the body biasing is applied to the body of $M_{N2}$. After applying \SI{0.3}{\volt}, the \textit{READ} current resolution is about \SI{165}{\nano\ampere}, which is 8.86x improved compared to the base test case. Here the max \textit{READ} power overhead is about 3.60\% compared to the base test case.
The \textit{READ} voltage on the source of the $M_{N2}$ is automatically adjusted based on the gate voltage of the $M_{N2}$. Due to that, the \textit{READ} power is quite stable in this design. Here the Monte Carlo simulation is observed with 1000 samples in a Cadence Spectre environment to analyze the \textit{READ} current variation. The \textit{READ} current shows about 0.65x variations with 8.68x resolution improvement. 


\begin{table*}[]
\centering
    \caption{\textit{READ} current resolution enhancement and design evaluation}
{%
\begin{tabular}{|l|l|l|l|l|l|}
\hline
Test Case &
  Device Size &
  \begin{tabular}[c]{@{}l@{}}READ current \\ Resolution\end{tabular} &
  \begin{tabular}[c]{@{}l@{}}READ current \\ Resolution  Evaluation\end{tabular} &
  \begin{tabular}[c]{@{}l@{}}Max READ \\ power  {[}both stage power{]}\end{tabular} &
  \begin{tabular}[c]{@{}l@{}}Max READ \\ power Evaluation\end{tabular} \\ \hline
Base sizing &
  \begin{tabular}[c]{@{}l@{}}MP1= (0.5/0.5)\SI{}{\micro\meter}   \\ MN1= (1/0.5)\SI{}{\micro\meter}   \\ MN2= (0.5/0.5)\SI{}{\micro\meter}\end{tabular} &
  \SI{19}{\nano\ampere} &
  - &
  \SI{18.87}{\micro\watt} &
  - \\ \hline
\begin{tabular}[c]{@{}l@{}}1st   stage \\ device sizing\end{tabular} &
  \begin{tabular}[c]{@{}l@{}}MP1= (1/0.5)\SI{}{\micro\meter}   \\ MN1= (4/0.5)\SI{}{\micro\meter}   \\ MN2= (0.5/0.5)\SI{}{\micro\meter}\end{tabular} &
  \SI{81}{\nano\ampere} &
  \begin{tabular}[c]{@{}l@{}}4.26x   improved \\ compared to \\ base sizing\end{tabular} &
  \SI{19.82}{\micro\watt} &
  \begin{tabular}[c]{@{}l@{}}5\% overhead \\ compared   to \\ base sizing\end{tabular} \\ \hline
\begin{tabular}[c]{@{}l@{}}2nd   state \\ device sizing\end{tabular} &
  \begin{tabular}[c]{@{}l@{}}MP1= (1/0.5)\SI{}{\micro\meter}   \\ MN1= (4/0.5)\SI{}{\micro\meter}   \\ MN2= (4/4)\SI{}{\micro\meter}\end{tabular} &
  \SI{98}{\nano\ampere} &
  \begin{tabular}[c]{@{}l@{}}5.16x   improved \\ compared to \\ base sizing\end{tabular} &
  \SI{18.84}{\micro\watt} &
  \begin{tabular}[c]{@{}l@{}}0.16\% improved   \\ compared to \\ base sizing\end{tabular} \\ \hline
\begin{tabular}[c]{@{}l@{}}V\_READ \\ @\SI{0.63}{\volt}\end{tabular} &
  \begin{tabular}[c]{@{}l@{}}MP1= (1/0.5)\SI{}{\micro\meter}   \\ MN1= (4/0.5)\SI{}{\micro\meter}   \\ MN2= (4/4)\SI{}{\micro\meter}\end{tabular} &
  \SI{143}{\nano\ampere} &
  \begin{tabular}[c]{@{}l@{}}7.53x improved   \\ compared to \\ base sizing\end{tabular} &
  \SI{19.6}{\micro\watt} &
  \begin{tabular}[c]{@{}l@{}}1.43\% overhead   \\ compared to \\ base sizing\end{tabular} \\ \hline
\begin{tabular}[c]{@{}l@{}}Body bias   \\ @\SI{0.3}{\volt}\end{tabular} &
  \begin{tabular}[c]{@{}l@{}}MP1= (1/0.5)\SI{}{\micro\meter}   \\ MN1= (4/0.5)\SI{}{\micro\meter}   \\ MN2= (4/4)\SI{}{\micro\meter}\end{tabular} &
  \SI{165}{\nano\ampere} &
  \begin{tabular}[c]{@{}l@{}}8.68x improved   \\ compared to \\ base sizing\end{tabular} &
  \SI{18.19}{\micro\watt} &
  \begin{tabular}[c]{@{}l@{}}3.60\% improved \\ compared   to \\ base sizing\end{tabular} \\ \hline
\end{tabular}%
}
\label{tab:table2}
\end{table*}


\section{Impact of READ Current resolution on Applications}

To evaluate the impact of READ current resolution on applications, we specifically wanted to evaluate how what impact READ errors would have on performance. Our hypothesis is that when the current resolution is higher, there will be fewer or no read errors, while when the current resolution is lower, read errors are significantly more likely. To perform this evaluation, we leveraged the TENNLab neuromorphic computing framework~\cite{plank2018tennlab}, which allows for evaluation of neuromorphic processors using different applications and algorithms.  Within the framework, we used the RISP neuromorphic simulator~\cite{plank2022case}, with integrate and fire neurons and synapses with 4-bit weight resolution.  

\begin{figure}
    \centering
    \includegraphics[width=0.5\textwidth]{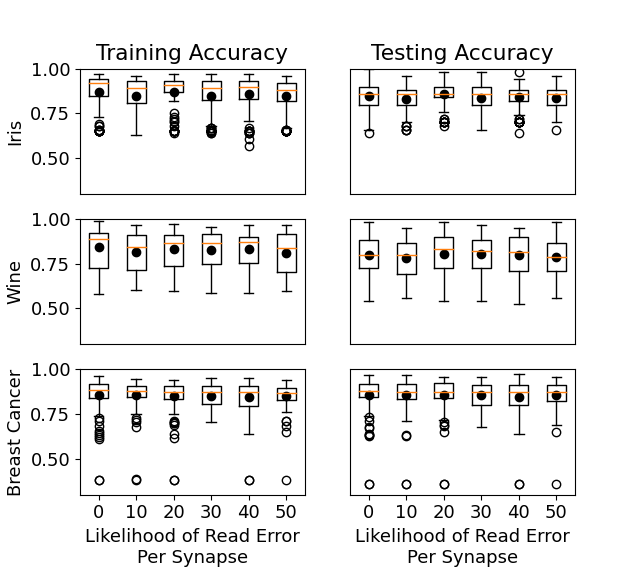}

    \caption{Training and testing accuracy for iris, wine, and breast cancer datasets.  In these results, the networks were trained and tested with read errors.}
    \label{fig:train_and_test_accuracy}
\end{figure}

\begin{figure}
    \centering
    \includegraphics[width=0.5\textwidth]{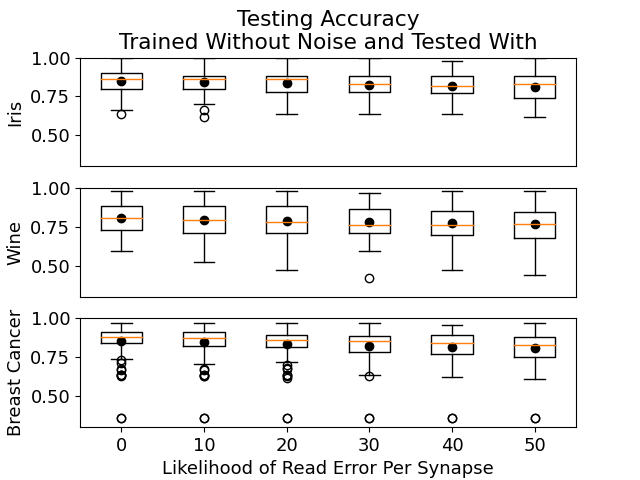}
    \caption{Testing accuracies for networks trained without read errors and tested with read errors.}
    \label{fig:train_without_test_with}
\end{figure}

To specifically study the impact of READ current resolution, we evaluated how spiking neural networks with different likelihoods of read failures for each synapse.  In particular, for a particular network evaluation, we defined a likelihood of read failures for each synapse read, wherein the weight read would be one level off (either the level above or level below), which would be more likely to happen for low current resolution. We trained the networks with these read failures using an evolutionary optimization training approach for spiking neural networks and neuromorphic system called EONS~\cite{schuman2020evolutionary}.  We trained for the iris dataset, the wine dataset, and the breast cancer dataset, three commonly used toy datasets in machine learning that are available in the UCI machine learning repository~\cite{asuncion2007uci}. We trained 100 networks for each of six different likelihoods of read errors for the synaptic weight values: 0, 10, 20, 30, 40, and 50 percent for each dataset. 
 Figure~\ref{fig:train_and_test_accuracy} shows the results for these simple datasets on both training and testing performance.  This figure shows that, in general, the best overall testing performance was achieved by no read errors at all; however, on average, some noise on the read errors does not necessarily hurt performance significantly, either in training or testing. For the results in Figure~\ref{fig:train_and_test_accuracy}, it is worth noting that the networks were trained and tested using read errors.  Figure~\ref{fig:train_without_test_with} shows the results for when networks are trained without read errors and then tested with varying likelihoods of read errors per synapse, which would likely be the case for networks trained in simulation and then deployed to hardware.  In this case, we can see that read errors cause a decrease in testing performance for each dataset.

\begin{figure}
        \centering
    \includegraphics[width=0.5\textwidth]{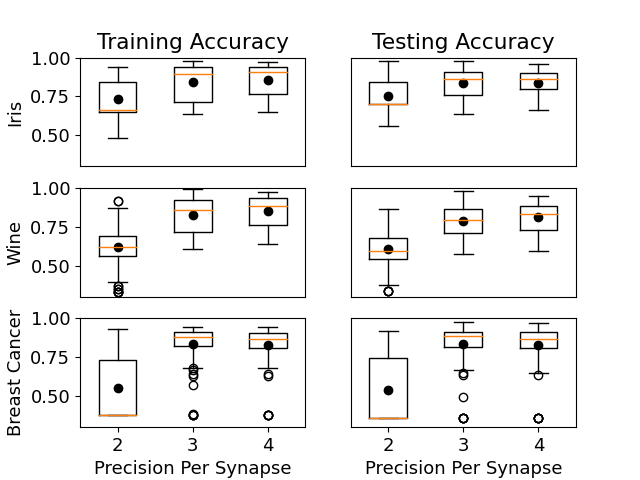}
    \caption{Results for reduced precision on the synaptic weights.}
    \label{fig:precision_impacts}
\end{figure}

Finally, because this work enables higher-precision weights to be used on synapses with more reliability, we investigated the impact of precision on performance.  Figure~\ref{fig:precision_impacts} shows the results for different levels of bit-precision (2, 3, and 4) on the synaptic weights. Here, we can see that 2-bit weight synapses perform significantly worse across all three datasets than 3- and 4-bit precision synapses, as expected.

\section{Comparison With Prior Work}  

\textit{READ} current resolution is a big challenge for the memristive-based synapse or memory design. Usually, the minimum current resolution is a few \SI{}{\nano\ampere}. Due to that, it is challenging for the circuit designer to sense the current level properly with ADC or CMOS neurons. In our neural network analysis, we consider a CMOS neuron to observe the charge accumulation and fire\cite{chakma}. In this work, a memristor-based synapse is optimized to enhance the \textit{READ} current resolutions. At first device sizing is considered to optimize the resolution. Hence, \textit{READ} voltage and body bias are considered to enhance the resolution at run time.  Here, the optimal size is adopted from TABLE \ref{tab:sizing}. In addition, the \textit{READ} voltage and body bias voltages are selected as \SI{0.63}{\volt} and \SI{0.3}{\volt} respectively. According to TABLE \ref{tab:com}, the minimum \textit{READ} current resolution is \SI{165}{\nano\ampere} with 4-bit data precision. As we know, the larger resistance level shows a lower current resolution. In our case, the current level difference between \SI{19}{\kilo\ohm} and \SI{20}{\kilo\ohm} is considered to determine the minimum current resolution. At room temperature the value is at least \SI{165}{\nano\ampere}. The maximum \textit{READ} current $\sim$ \SI{28}{\nano\ampere} is captured at \SI{5}{\kilo\ohm}. Our synapse is programmed at LRS to avoid inherent process variations. 

\begin{table*}[ht]
\centering
    \caption{Comparison with prior works}
\resizebox{\textwidth}{!}
{%
\begin{tabular}{|l|l|l|l|l|l|l|l|l|l|}
\hline
Reference &
  \cite{r1} &
  \cite{r2} &
  \cite{r3} &
  \cite{r4} &
  \cite{r5} &
  \cite{Amorphous} &
  \cite{r7} &
  \cite{r8} &
  This work \\ \hline
Technology &
  \SI{65}{\nano\meter} CMOS &
  \SI{65}{\nano\meter} CMOS &
  - &
  - &
  - &
  - &
  - &
   - &
   \SI{65}{\nano\meter} CMOS \\ \hline
\begin{tabular}[c]{@{}l@{}}Memristor \\ Material\end{tabular} &
  \ce{HfO2} &
  \ce{HfO2} &
  \ce{TiO2} &
  \ce{TiO2} &
  \ce{HfO2} &
  \ce{a-ZnO} &
  \ce{HfO2} &
  \ce{TiO2} &
  \ce{HfO2} \\ \hline
\begin{tabular}[c]{@{}l@{}}Programming \\ Reason\end{tabular} &
  \SI{5}{\kilo\ohm} - \SI{20}{\kilo\ohm} & 
  \SI{3}{\kilo\ohm} - \SI{18}{\kilo\ohm} &
  \SI{1}{\kilo\ohm} - \SI{1}{\giga\ohm} &
  \SI{10}{\kilo\ohm} - \SI{100}{\mega\ohm} &
  \SI{10}{\kilo\ohm} - \SI{1}{\mega\ohm} &
  \SI{24}{\mega\ohm} - \SI{176}{\mega\ohm} &
  \SI{0.8}{\kilo\ohm} - \SI{100}{\mega\ohm} &
  \SI{1}{\mega\ohm} - \SI{10}{\giga\ohm} &
 \SI{5}{\kilo\ohm} - \SI{20}{\kilo\ohm} \\ \hline
\begin{tabular}[c]{@{}l@{}}Impact of \\ Process\\ Variation \end{tabular} &
  Lower &
  Lower &
  Higher &
  Higher &
  Higher &
  Higher &
  Higher &
  Higher &
  Lower \\ \hline
\begin{tabular}[c]{@{}l@{}}Number of \\ Programming  \\ States\end{tabular} &
  16 &
  16 &
  4 &
  6 &
  4 &
  5 &
  8 &
  6 &
  16 \\ \hline
\begin{tabular}[c]{@{}l@{}}Storage \\ Density \\ Improvement \\compared to\\ prior work \end{tabular} &
  Same &
  Same &
  4x &
  2.67x &
  4x &
  3.2x &
  2x &
  2.67x &
  - \\ \hline
\begin{tabular}[c]{@{}l@{}}Max READ \\ Current\end{tabular} &
  $\sim$\SI{5.44}{\micro\ampere} &
  $\sim$\SI{2.92}{\micro\ampere} &
  $\sim$\SI{90}{\micro\ampere} &
 \SI{100}{\micro\ampere} &
 \SI{10}{\micro\ampere} &
  \SI{4.166}{\nano\ampere} &
  $\sim$\SI{625}{\micro\ampere} &
  \SI{0.5}{\micro\ampere} &
  \SI{28}{\micro\ampere}\\ \hline
\begin{tabular}[c]{@{}l@{}}Max READ \\ Power\end{tabular} &
 $\sim$\SI{8.24}{\micro\watt} &
 $\sim$\SI{3.50}{\micro\watt} &
  $\sim$\SI{45}{\micro\watt} &
 \SI{100}{\micro\watt} &
 \SI{1}{\micro\watt} &
  \SI{0.4116}{\nano\watt} &
  $\sim$\SI{312.5}{\micro\watt} &
  \SI{0.25}{\micro\watt} &
  \SI{19.6}{\micro\watt}\\ \hline
  \begin{tabular}[c]{@{}l@{}}Max READ \\ Power \\ Improvement \\compared to\\ prior work\end{tabular} &
 No &
 No &
  Yes &
 Yes &
 No &
  No &
  Yes &
  No &
  -\\ \hline
\begin{tabular}[c]{@{}l@{}}Minimum READ  \\ Current Resolution\end{tabular} &
  \SI{20}{\nano\ampere} &
  \SI{20}{\nano\ampere} &
  \SI{90}{\nano\ampere} &
 $\sim$\SI{17}{\nano\ampere} &
 \SI{81}{\nano\ampere} &
 \SI{0.462}{\nano\ampere}&
  \SI{45}{\nano\ampere} &
  \SI{0.45}{\nano\ampere} &
  \SI{165}{\nano\ampere} \\ \hline
\begin{tabular}[c]{@{}l@{}}READ Current \\ Resolution \\ Improvement \\compared to\\ prior work\end{tabular} &
  8.25x &
  8.25x &
  1.83x &
  $\sim$9.7x &
  2.04x &
  357x &
  3.67x &
  366.67x &
  - \\ \hline
  \begin{tabular}[c]{@{}l@{}}READ Current \\ Linearity\end{tabular} &
  \begin{tabular}[c]{@{}l@{}}Linear\end{tabular} &
  \begin{tabular}[c]{@{}l@{}}Linear\end{tabular} &
  \begin{tabular}[c]{@{}l@{}}Non-linear\end{tabular} &
  \begin{tabular}[c]{@{}l@{}}Non-linear\end{tabular} &
  \begin{tabular}[c]{@{}l@{}}Non-linear\end{tabular} &
  \begin{tabular}[c]{@{}l@{}}Non-linear\end{tabular} &
  \begin{tabular}[c]{@{}l@{}}Non-linear\end{tabular} &
  \begin{tabular}[c]{@{}l@{}}Non-linear\end{tabular} &
  \begin{tabular}[c]{@{}l@{}}Linear\end{tabular} \\ \hline
\end{tabular}%
}
\label{tab:com}
\end{table*}

Another research article shows the minimum \textit{READ} current resolution of a 3T1R synapse is \SI{20}{\nano\ampere} for 4-bit data precision. Our proposed design shows 8.25x resolution improvement compared to their work\cite{r1}. Energy-efficient and high-performance synapse is presented in another research paper, where the minimum \textit{READ} current resolution is \SI{20}{\nano\ampere} \cite{r2}. This design is also based on \ce{HfO2} based memristor and \SI{65}{\nano\meter} CMOS process. The programming region is between \SI{3}{\kilo\ohm} to \SI{18}{\kilo\ohm}, which provides low inherent process variation with 4-bit data density. Our proposed design also illustrated 8.25x current resolution improvement compared to their work \cite{r2}.  A \ce{TiO2} based memristive memory is presented in \cite{r3}, which is programmed from \SI{1}{\kilo\ohm} to \SI{1}{\giga\ohm}. Due to the utilization of HRS, the inherent process variation is higher for this design compare to our proposed design. This design also covers four programming states. As a result, our proposed design shows a 4x more dense data storage capacity compared to their design \cite{r3}. The max \textit{READ} current is about 3.21x higher than our proposed design. The minimum \textit{READ}  current resolution is about \SI{90}{\nano\ampere}, which is 1.83x lower than our proposed design. 

Another research group is presented a \ce{TiO2} based multi-level resistive memory, which is programmed from \SI{10}{\kilo\ohm} to \SI{100}{\mega\ohm} \cite{r4}. Due to programming in HRS, the inherent process variation is higher than our design. Their design is only programmed in six different states, which exhibits 2.67x lower memory density compared to our proposed design. The maximum \textit{READ} current of their design is \SI{100}{\micro\ampere}, which is 3.57x higher than our proposed design. Moreover, the minimum \textit{READ} current resolution of their design is $\sim$ \SI{17}{\nano\ampere}, whereas our proposed design shows $\sim$ 9.7x enhanced resolution compared to their design \cite{r4}. A \ce{HfO2} based multi-level cell is presented with 2-bit memory density, which is 4x  lower memory density than our proposed synapse \cite{r5}. Their device is programmed from \SI{10}{\kilo\ohm} to \SI{1}{\mega\ohm}, which shows higher inherent process variation compared to our design. Their device draws lower maximum \textit{READ} current than our design. However, the minimum \textit{READ} current resolution is 2.04x lower than our proposed design.

Another multi-level resistive memory is presented using \ce{a-ZnO}, which is programmed between  \SI{24}{\mega\ohm} to \SI{176}{\mega\ohm}\cite{Amorphous}. Their design shows higher process variation and 3.2x lower memory density than our proposed design. Their maximum \textit{READ} current is lower than our design. However, the minimum \textit{READ} current resolution is extremely low compared to our design.   A \ce{HfO2} based multi-level RRAM is presented in a research article, where authors programmed their device between \SI{0.8}{\kilo\ohm} to \SI{100}{\mega\ohm} with 8 programming states \cite{r7}. Our proposed design shows lower process variation and 2x higher memory density than their design. The maximum \textit{READ} current of their design is extremely higher than our proposed design. Our proposed design also shows a 3.67x improved minimum \textit{READ} current resolution than their design\cite{r7}. 
Another research group presented their device with \ce{TiO2}, which is programmed between \SI{1}{\mega\ohm} to \SI{10}{\giga\ohm} with six programming states\cite{r8}. This design shows a 2.67x lower memory density than our proposed design. Due to their HRS programming region, the process variation is higher than our design. The maximum \textit{READ} current of their design is lower than our proposed design. However, their minimum \textit{READ} current resolution is extremely lower than our proposed design. Moreover, our design shows power savings compare to prior works \cite{r3,r4,r7}. In addition, our proposed design shows more linear \textit{READ} current compared to other designs \cite{r3,r4,r5,Amorphous,r7,r8}. Due to a compact programming range, our design shows better linearity compared to prior works. A wide programming range causes non-linear behavior of \textit{READ} current.    

After comparing our proposed design with prior works from different research groups, it can be said that our proposed design shows lower inherent process variation with higher memory density. In addition, our proposed design shows an enhanced \textit{READ} current resolution compared to others' designs.

\section{Conclusions and future work}
In this paper, a \ce{HfO2} based current-controlled memristive synapse is optimized for \textit{READ} operation. At first, the \textit{READ} devices are optimized to enhance the \textit{READ} current resolution. About 4.3x and 21\% READ current resolution is enhanced with $1^{st}$ and  $2^{nd}$ stage device sizing respectively. \textit{READ} voltage scaling and body biasing are applied to enhance the \textit{READ} current resolution at run time. About 46\% and 15\% \textit{READ} current resolution is improved with \textit{READ} voltage scaling and body biasing.
A neuromorphic framework shows that a higher \textit{READ} current resolution exhibits better accuracy compared to a lower resolution on a simple classification application. Lower resolutions are more likely to be affected by reading failures with higher noise. As a result, a higher \textit{READ} current resolution makes the neuromorphic system more reliable.      
  
\section*{Acknowledgments}
The authors are thankful to Maximilian Liehr and Rocco Febbo for their suggestions and help in device testing.  
    
    \bibliographystyle{IEEEtran}
    \bibliography{bibliography}

\begin{thebibliography}{10}
\providecommand{\url}[1]{#1}
\csname url@samestyle\endcsname
\providecommand{\newblock}{\relax}
\providecommand{\bibinfo}[2]{#2}
\providecommand{\BIBentrySTDinterwordspacing}{\spaceskip=0pt\relax}
\providecommand{\BIBentryALTinterwordstretchfactor}{4}
\providecommand{\BIBentryALTinterwordspacing}{\spaceskip=\fontdimen2\font plus
\BIBentryALTinterwordstretchfactor\fontdimen3\font minus
  \fontdimen4\font\relax}
\providecommand{\BIBforeignlanguage}[2]{{%
\expandafter\ifx\csname l@#1\endcsname\relax
\typeout{** WARNING: IEEEtran.bst: No hyphenation pattern has been}%
\typeout{** loaded for the language `#1'. Using the pattern for}%
\typeout{** the default language instead.}%
\else
\language=\csname l@#1\endcsname
\fi
#2}}
\providecommand{\BIBdecl}{\relax}
\BIBdecl

\bibitem{intro1}
\BIBentryALTinterwordspacing
A.~Krizhevsky, I.~Sutskever, and G.~E. Hinton, ``Imagenet classification with
  deep convolutional neural networks,'' in \emph{Advances in Neural Information
  Processing Systems}, F.~Pereira, C.~Burges, L.~Bottou, and K.~Weinberger,
  Eds., vol.~25.\hskip 1em plus 0.5em minus 0.4em\relax Curran Associates,
  Inc., 2012. [Online]. Available:
  \url{https://proceedings.neurips.cc/paper_files/paper/2012/file/c399862d3b9d6b76c8436e924a68c45b-Paper.pdf}
\BIBentrySTDinterwordspacing

\bibitem{jetc}
\BIBentryALTinterwordspacing
M.~M. Adnan, S.~Sayyaparaju, S.~D. Brown, M.~S.~A. Shawkat, C.~D. Schuman, and
  G.~S. Rose, ``Design of a robust memristive spiking neuromorphic system with
  unsupervised learning in hardware,'' \emph{J. Emerg. Technol. Comput. Syst.},
  vol.~17, no.~4, jun 2021. [Online]. Available:
  \url{https://doi.org/10.1145/3451210}
\BIBentrySTDinterwordspacing

\bibitem{intro2}
K.~Lee, J.~Park, and H.-J. Yoo, ``A low-power, mixed-mode neural network
  classifier for robust scene classification,'' \emph{JOURNAL OF SEMICONDUCTOR
  TECHNOLOGY AND SCIENCE}, vol.~19, pp. 129--136, 02 2019.

\bibitem{intro3}
S.~Kim, H.~Kim, S.~Hwang, M.-H. Kim, Y.-F. Chang, and B.-G. Park, ``Analog
  synaptic behavior of a silicon nitride memristor,'' \emph{ACS Applied
  Materials \& Interfaces}, vol.~9, 10 2017.

\bibitem{intro4}
P.~Rashvand, M.~Ahmadzadeh, and F.~Shayegh, ``Design and implementation of a
  spiking neural network with integrate-and-fire neuron model for pattern
  recognition,'' \emph{International Journal of Neural Systems}, vol.~31, p.
  2050073, 12 2020.

\bibitem{RC_glsvlsi}
\BIBentryALTinterwordspacing
M.~Rathore, R.~Febbo, A.~Foshie, S.~N.~B. Tushar, H.~Das, and G.~S. Rose,
  ``Reliability analysis of memristive reservoir computing architecture,'' in
  \emph{Proceedings of the Great Lakes Symposium on VLSI 2023}, ser. GLSVLSI
  '23.\hskip 1em plus 0.5em minus 0.4em\relax New York, NY, USA: Association
  for Computing Machinery, 2023, p. 131–136. [Online]. Available:
  \url{https://doi.org/10.1145/3583781.3590210}
\BIBentrySTDinterwordspacing

\bibitem{mer}
P.~Merolla, J.~Arthur, F.~Akopyan, N.~Imam, R.~Manohar, and D.~S. Modha, ``A
  digital neurosynaptic core using embedded crossbar memory with 45pj per spike
  in 45nm,'' in \emph{2011 IEEE Custom Integrated Circuits Conference (CICC)},
  2011, pp. 1--4.

\bibitem{sensors}
M.~Asghar, S.~Arslan, and H.~Kim, ``A low-power spiking neural network chip
  based on a compact lif neuron and binary exponential charge injector synapse
  circuits,'' \emph{Sensors}, vol.~21, p. 4462, 06 2021.

\bibitem{comp}
\BIBentryALTinterwordspacing
Y.~Kim, Y.~Zhang, and P.~Li, ``A reconfigurable digital neuromorphic processor
  with memristive synaptic crossbar for cognitive computing,'' \emph{J. Emerg.
  Technol. Comput. Syst.}, vol.~11, no.~4, apr 2015. [Online]. Available:
  \url{https://doi.org/10.1145/2700234}
\BIBentrySTDinterwordspacing

\bibitem{adam_glsvlsi}
\BIBentryALTinterwordspacing
A.~Z. Foshie, C.~Rizzo, H.~Das, C.~Zheng, J.~S. Plank, and G.~S. Rose,
  ``Benchmark comparisons of spike-based reconfigurable neuroprocessor
  architectures for control applications,'' in \emph{Proceedings of the Great
  Lakes Symposium on VLSI 2022}, ser. GLSVLSI '22.\hskip 1em plus 0.5em minus
  0.4em\relax New York, NY, USA: Association for Computing Machinery, 2022, p.
  383–386. [Online]. Available: \url{https://doi.org/10.1145/3526241.3530381}
\BIBentrySTDinterwordspacing

\bibitem{adam}
A.~Z. Foshie, N.~N. Chakraborty, J.~J. Murray, T.~J. Fowler, M.~S. Ara~Shawkat,
  and G.~S. Rose, ``A multi-context neural core design for reconfigurable
  neuromorphic arrays,'' in \emph{2021 IEEE Computer Society Annual Symposium
  on VLSI (ISVLSI)}, 2021, pp. 67--72.

\bibitem{glsvlsi2023}
\BIBentryALTinterwordspacing
N.~N. Chakraborty, H.~Das, and G.~S. Rose, ``A mixed-signal short-term
  plasticity implementation for a current-controlled memristive synapse,'' in
  \emph{Proceedings of the Great Lakes Symposium on VLSI 2023}, ser. GLSVLSI
  '23.\hskip 1em plus 0.5em minus 0.4em\relax New York, NY, USA: Association
  for Computing Machinery, 2023, p. 179–182. [Online]. Available:
  \url{https://doi.org/10.1145/3583781.3590283}
\BIBentrySTDinterwordspacing

\bibitem{STDP}
R.~Weiss, H.~Das, N.~N. Chakraborty, and G.~S. Rose, ``Stdp based online
  learning for a current-controlled memristive synapse,'' in \emph{2022 IEEE
  65th International Midwest Symposium on Circuits and Systems (MWSCAS)}, 2022,
  pp. 1--4.

\bibitem{Cruz}
\BIBentryALTinterwordspacing
J.~M. Cruz-Albrecht, T.~Derosier, and N.~Srinivasa, ``A scalable neural chip
  with synaptic electronics using cmos integrated memristors,''
  \emph{Nanotechnology}, vol.~24, no.~38, p. 384011, sep 2013. [Online].
  Available: \url{https://dx.doi.org/10.1088/0957-4484/24/38/384011}
\BIBentrySTDinterwordspacing

\bibitem{brivio}
S.~Brivio, D.~Conti, M.~Nair, J.~Frascaroli, E.~Covi, C.~Ricciardi,
  G.~Indiveri, and S.~Spiga, ``Extended memory lifetime in spiking neural
  networks employing memristive synapses with nonlinear conductance dynamics,''
  \emph{Nanotechnology}, vol.~30, 10 2018.

\bibitem{memr}
L.~Chua, ``Memristor-the missing circuit element,'' \emph{IEEE Transactions on
  Circuit Theory}, vol.~18, no.~5, pp. 507--519, 1971.

\bibitem{ref_tab3}
\BIBentryALTinterwordspacing
Z.~I. Mannan, H.~Kim, and L.~Chua, ``Implementation of neuro-memristive synapse
  for long-and short-term bio-synaptic plasticity,'' \emph{Sensors}, vol.~21,
  no.~2, p. 644, Jan 2021. [Online]. Available:
  \url{http://dx.doi.org/10.3390/s21020644}
\BIBentrySTDinterwordspacing

\bibitem{r6}
M.~Payvand, Y.~Demirag, T.~Dalgaty, E.~Vianello, and G.~Indiveri, ``Analog
  weight updates with compliance current modulation of binary rerams for
  on-chip learning,'' in \emph{2020 IEEE International Symposium on Circuits
  and Systems (ISCAS)}, 2020, pp. 1--5.

\bibitem{r2}
N.~N. Chakraborty, H.~Das, and G.~S. Rose, ``Energy efficient and
  high-performance synaptic operating point evaluation for snn applications,''
  in \emph{2023 IEEE 66th International Midwest Symposium on Circuits and
  Systems (MWSCAS)}, 2023.

\bibitem{model_glsvlsi}
H.~Das, M.~Rathore, R.~Febbo, M.~Liehr, N.~C. Cady, and G.~S. Rose, ``Rfam:
  Reset-failure-aware-model for hfo2-based memristor to enhance the reliability
  of neuromorphic design,'' in \emph{Proceedings of the Great Lakes Symposium
  on VLSI 2023}, 2023, pp. 281--286.

\bibitem{r1}
H.~Das, R.~D. Febbo, C.~Rizzo, N.~N. Chakraborty, J.~S. Plank, and G.~S. Rose,
  ``Optimizations for a current-controlled memristor-based neuromorphic synapse
  design,'' \emph{arXiv preprint arXiv:2305.16418}, 2023.

\bibitem{plank2018tennlab}
J.~S. Plank, C.~D. Schuman, G.~Bruer, M.~E. Dean, and G.~S. Rose, ``The tennlab
  exploratory neuromorphic computing framework,'' \emph{IEEE Letters of the
  Computer Society}, vol.~1, no.~2, pp. 17--20, 2018.

\bibitem{plank2022case}
J.~S. Plank, C.~Zheng, B.~Gullett, N.~Skuda, C.~Rizzo, C.~D. Schuman, and G.~S.
  Rose, ``The case for risp: A reduced instruction spiking processor,''
  \emph{arXiv preprint arXiv:2206.14016}, 2022.

\bibitem{schuman2020evolutionary}
C.~D. Schuman, J.~P. Mitchell, R.~M. Patton, T.~E. Potok, and J.~S. Plank,
  ``Evolutionary optimization for neuromorphic systems,'' in \emph{Proceedings
  of the Neuro-inspired Computational Elements Workshop}, 2020, pp. 1--9.

\bibitem{asuncion2007uci}
A.~Asuncion and D.~Newman, ``Uci machine learning repository,'' 2007.

\bibitem{chakma}
G.~Chakma, M.~M. Adnan, A.~R. Wyer, R.~Weiss, C.~D. Schuman, and G.~S. Rose,
  ``Memristive mixed-signal neuromorphic systems: Energy-efficient learning at
  the circuit-level,'' \emph{IEEE Journal on Emerging and Selected Topics in
  Circuits and Systems}, vol.~8, no.~1, pp. 125--136, 2018.

\bibitem{r3}
M.~Terai, Y.~Sakotsubo, S.~Kotsuji, and H.~Hada, ``Resistance controllability
  of $\hbox{Ta}_{2} \hbox{O}_{5}/\hbox{TiO}_{2}$ stack reram for low-voltage
  and multilevel operation,'' \emph{IEEE Electron Device Letters}, vol.~31,
  no.~3, pp. 204--206, 2010.

\bibitem{r4}
P.~Bousoulas, I.~Giannopoulos, P.~Asenov, I.~Karageorgiou, and D.~Tsoukalas,
  ``Investigating the origins of high multilevel resistive switching in forming
  free ti/tio2- x-based memory devices through experiments and simulations,''
  \emph{Journal of Applied Physics}, vol. 121, no.~9, p. 094501, 2017.

\bibitem{r5}
W.~Chen, W.~Lu, B.~Long, Y.~Li, D.~Gilmer, G.~Bersuker, S.~Bhunia, and R.~Jha,
  ``Switching characteristics of w/zr/hfo2/tin reram devices for multi-level
  cell non-volatile memory applications,'' \emph{Semiconductor Science and
  Technology}, vol.~30, no.~7, p. 075002, 2015.

\bibitem{Amorphous}
Y.~Huang, Z.~Shen, Y.~Wu, X.~Wang, S.~Zhang, X.~Shi, and H.~Zeng, ``Amorphous
  zno based resistive random access memory,'' \emph{RSC advances}, vol.~6,
  no.~22, pp. 17\,867--17\,872, 2016.

\bibitem{r7}
H.-S. Philip{\'a}Wong \emph{et~al.}, ``Multi-level control of conductive
  nano-filament evolution in hfo 2 reram by pulse-train operations,''
  \emph{Nanoscale}, vol.~6, no.~11, pp. 5698--5702, 2014.

\bibitem{r8}
M.~Tsigkourakos, P.~Bousoulas, V.~Aslanidis, E.~Skotadis, and D.~Tsoukalas,
  ``Ultra-low power multilevel switching with enhanced uniformity in forming
  free tio2- x-based rram with embedded pt nanocrystals,'' \emph{physica status
  solidi (a)}, vol. 214, no.~12, p. 1700570, 2017.

\end{thebibliography}

\vspace{11pt}
\bf{}\vspace{-33pt}
\begin{IEEEbiography}[{\includegraphics[width=1in,height=1.25in,clip,keepaspectratio]{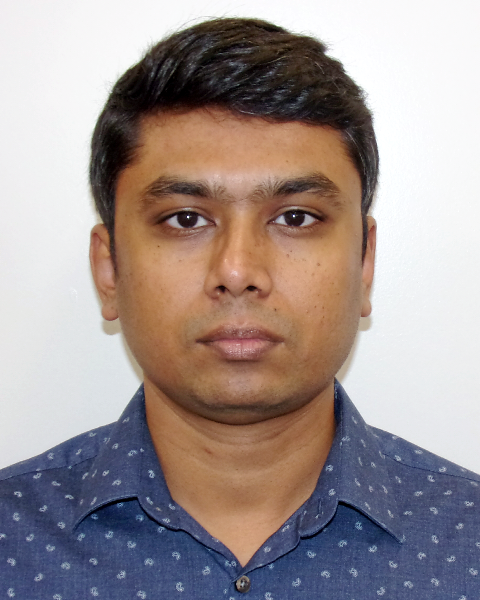}}]{Hritom Das} (Member, IEEE) received the B.Sc. degree in electrical and electronic engineering from American International University-Bangladesh, Dhaka, Bangladesh, the M.Sc. degree in electronic engineering from Kyungpook National University, Daegu, South Korea, and the Ph.D. degree.in electrical and computer engineering from North Dakota State University, Fargo, North Dakota, in 2012, 2015 and 2020 respectively. He was a visiting Assistant Professor with the Department of electrical and computer engineering at the University of South Alabama, Mobile, AL, USA. Currently he is a Post-Doctoral Research Associate with the department of electrical engineering and computer science at The University of Tennessee, Knoxville, TN, USA. His research interest includes low power VLSI circuit design, optimization, and testing. He is also exploring machine learning implementation on traditional electronics.  In addition, he is working on neuromorphic system design and optimization.
\end{IEEEbiography}

 \vskip 0pt plus -1fil
 
\begin{IEEEbiography}[{\includegraphics[width=1in,height=1.25in,clip,keepaspectratio]{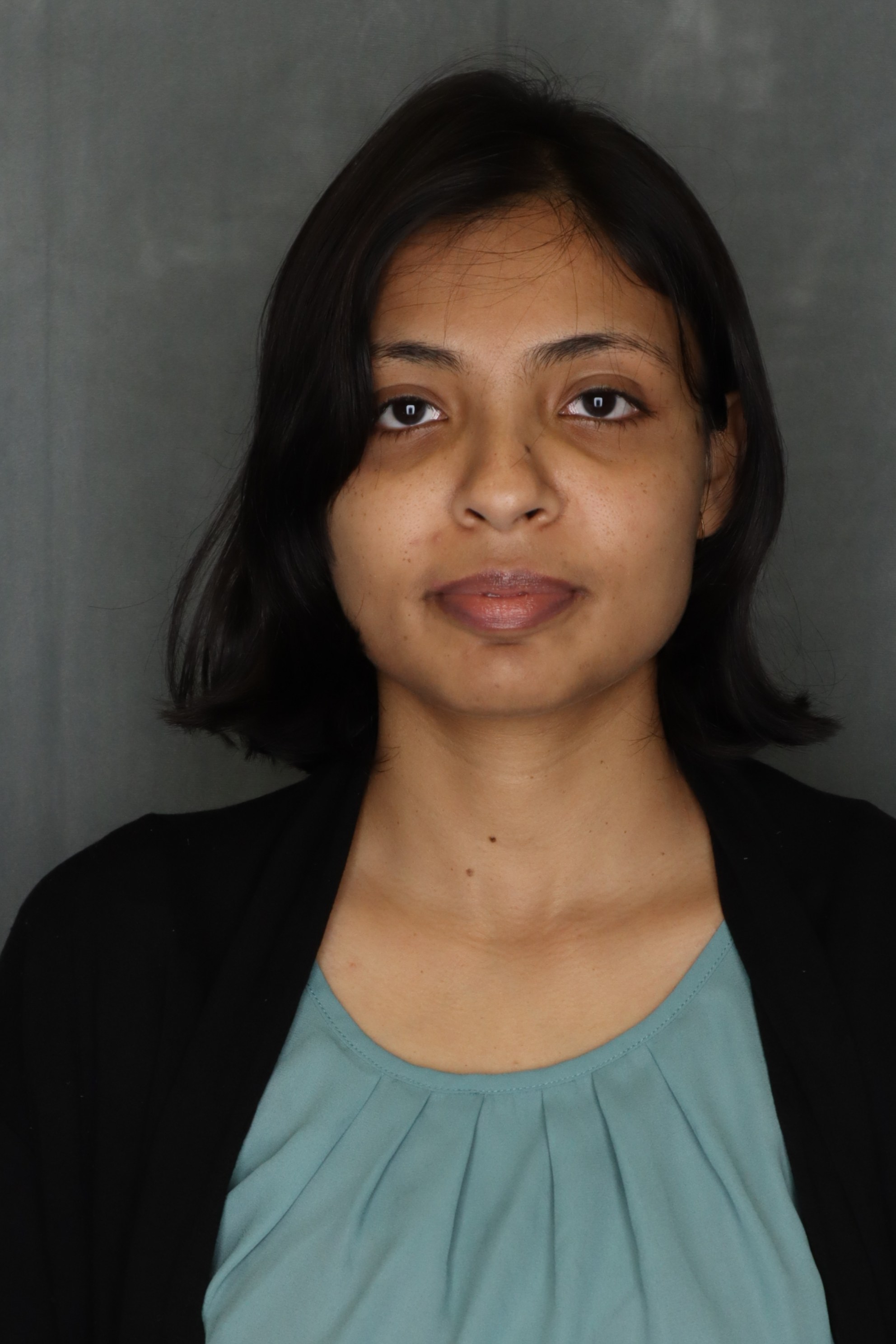}}] {Nishith N. Chakraborty} (Member, IEEE) is currently enrolled at the University of Tennessee, Knoxville (UTK) as a Ph.D student. She received her B.Sc. degree in electrical and electronic engineering from Bangladesh University of Engineering and Technology, and MS in electrical engineering from University of California, Riverside. Her research interest includes analog mixed-signal neuromorphic circuits design and optimization, low-power VLSI circuits and memristor-based circuit design.
\end{IEEEbiography}

\vskip 0pt plus -1fil

\begin{IEEEbiography}[{\includegraphics[width=1in,height=1.25in,clip,keepaspectratio]{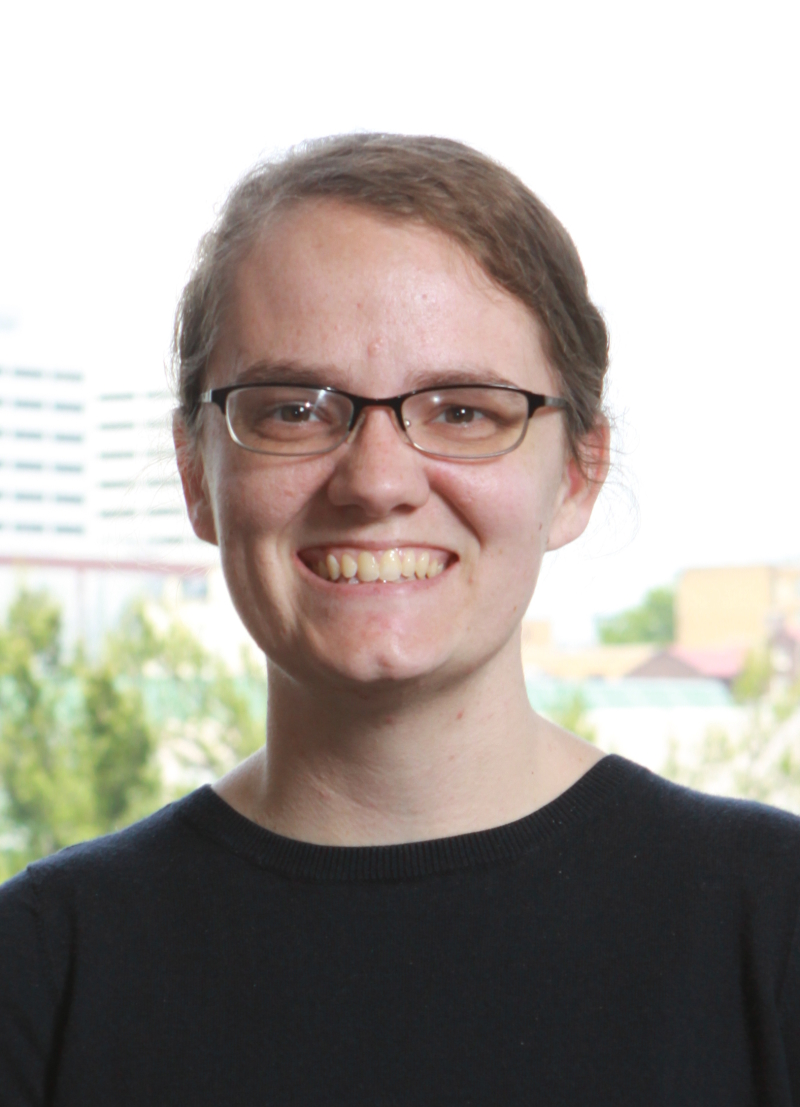}}] {Catherine (Katie) D. Schuman} (Senior Member, IEEE) received the Ph.D. degree in computer science from The University of Tennessee (UT), Knoxville, TN, USA, in 2015, where she completed her dissertation on the use of evolutionary algorithms to train spiking neural networks for neuromorphic systems. She was a Research Scientist at the Oak Ridge National Laboratory, Oak Ridge, TN, USA, where her research focused on algorithms and applications of neuromorphic systems. She is an Assistant Professor with the Department of Electrical Engineering and Computer Science, UT. She co-leads the TENNLab Neuromorphic Computing Research Group at UT. She has over 100 publications and holds eight patents in the field of neuromorphic computing.
Dr. Schuman received the Department of Energy Early Career Award in 2019.
\end{IEEEbiography}

\vskip 0pt plus -1fil

\begin{IEEEbiography}[{\includegraphics[width=1in,height=1.25in,clip,keepaspectratio]{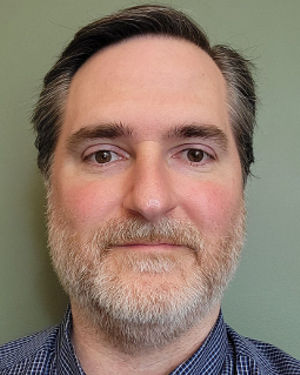}}] {Garrett S. Rose} (Senior Member, IEEE) received
the B.S. degree in computer engineering from Virginia Polytechnic Institute and State University (Virginia Tech), Blacksburg, VA, USA, in 2001, and the M.S. and Ph.D. degrees in electrical engineering from the University of Virginia, Charlottesville, VA, USA,in 2003 and 2006, respectively. His Ph.D. dissertation was on the topic of circuit design methodologies for molecular electronic circuits and computing architectures. He is currently a Professor with the Min H. Kao Department of Electrical Engineering and Computer Science, University of Tennessee, Knoxville, TN, USA, where his work is focused on research in the areas of nanoelectronic circuit design, neuromorphic computing, and hardware security. From June 2011 to July 2014, he was with the Air Force Research Laboratory, Information Directorate, Rome, NY, USA. From August 2006 to May 2011, he was an Assistant Professor with the Department of Electrical and Computer Engineering, Polytechnic Institute of New York University, Brooklyn, NY, USA. From May 2004 to August 2005, he was with MITRE Corporation, McLean, VA, USA, involved in the design and simulation of nanoscale circuits and systems. His research interests include low-power circuits, system-on-chip design, trusted hardware, and developing VLSI design methodologies for novel nanoelectronic technologies.
\end{IEEEbiography}

\end{document}